\documentclass[%
reprint,
superscriptaddress,
 amsmath,amssymb,
 aps,
floatfix,
]{revtex4-2}

\usepackage{graphicx}
\usepackage{dcolumn}
\usepackage{bm}
\usepackage{appendix}

\usepackage{color}
\usepackage{physics}

\DeclareRobustCommand{\erase}{\bgroup\markoverwith{\textcolor{red}{\rule[.5ex]{2pt}{0.4pt}}}\ULon}

\begin{document}
\preprint{APS/123-QED}

\title{Surface-Normal Electric Spin in Obliquely Excited Parallel-Slit SPP Fields}

\author{Naoki Ichiji}
\affiliation{National Institute for Materials Science, 1-1 Namiki, Tsukuba, Ibaraki 305-0044, Japan}
\affiliation{Institute of Industrial Science, The University of Tokyo, 4-6-1 Komaba, Meguro-Ku, Tokyo 153-8505, Japan}

\begin{abstract}
Handedness-dependent light–matter interactions require control of local electric-field rotation, but generating controlled surface-normal spin textures is difficult under oblique illumination. Here, we show that oblique $s$-polarized illumination of a periodic array of parallel slits produces spatially alternating surface-normal electric spin, with its period tunable by the incidence angle. We further show that the spin texture changes qualitatively between substrate- and air-side excitation depending on whether the SPP field interferes with the excitation field. Under substrate-side excitation, the spin distribution follows the conventional spin-flow relation, whereas under air-side excitation, interference between the SPP and excitation fields produces a distinct spatial periodicity and allows the spin extrema to coincide with the field-intensity maxima. A simple interference model reproduces these contrasting spin textures.
\end{abstract}

\maketitle
Light--matter interactions are governed not only by optical intensity but also by the polarization state of light. In particular, material responses sensitive to the handedness of light have gained renewed attention, including circular-polarization-resolved Raman scattering~\cite{Ishito23NatP}, valley-selective excitation in two-dimensional semiconductors~\cite{Mak12NatN}, and magneto-optical or optomagnetic responses~\cite{Choi17NatCom}.

Among the quantities used to describe optical handedness, electric spin characterizes the magnitude and direction of local electric-field rotation. Spatial variations in the magnitude and three-dimensional orientation of this vector form optical spin textures, which have attracted growing interest. For a surface or two-dimensional material lying in the \(xy\) plane, the surface-normal component of electric spin is particularly relevant, as it characterizes the rotation of the in-plane field. This field can be decomposed into the circular-basis components
\begin{equation}
E_{\pm}=\frac{E_x\pm iE_y}{\sqrt{2}}.
\end{equation}
The corresponding intensity imbalance,
$\Delta_\sigma=|E_+|^2-|E_-|^2 = -2\operatorname{Im}(E_x^*E_y)$, quantifies the handedness of
this temporal rotation and is proportional to the surface-normal component of the electric spin density~\cite{Neugebauer18PRX, Forbes24JO}.

\begin{figure}[b!]
\centering
\includegraphics[width=0.5\textwidth]{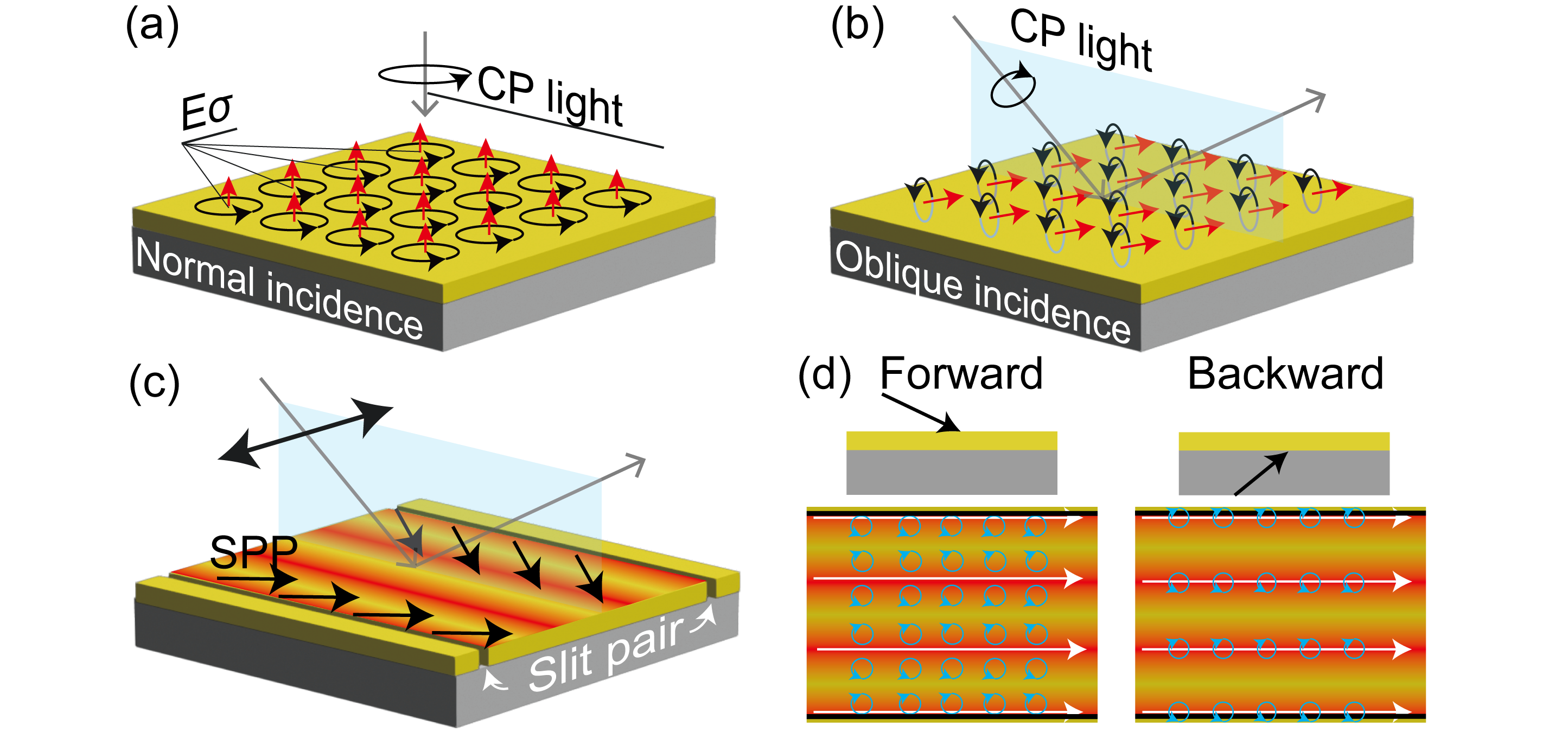}
\caption{Surface-normal electric spin under normal and oblique excitation. (a,b) Circularly polarized (CP) light at (a) normal and (b) oblique incidence. (c) Two-slit SPP interference excited by oblique $s$-polarized light. (d) Field intensity and in-plane field rotation for forward air-side (left)
and backward substrate-side (right) excitation. White arrows indicate energy flow, while light-blue circular arrows indicate the handedness of local electric-field rotation.}
\label{Fig:concept}
\end{figure}

At normal incidence, the circularly rotating incident electric field projects directly onto the sample plane (Fig.~\ref{Fig:concept}(a)). At oblique incidence, however, geometrical projection and the distinct complex Fresnel responses of the $s$- and $p$-polarized components alter their relative in-plane amplitudes and phases, so the incident circular polarization is generally not preserved at the surface (Fig.~\ref{Fig:concept}(b)). In practice, many surface-sensitive measurements employ oblique illumination. In PEEM, optical access is often constrained by the electron-optical geometry~\cite{Word17Ultra}, while pump--probe measurements may employ noncollinear incidence to spatially separate the reflected pump and probe beams before detection~\cite{Qin20PR}. These constraints make it challenging to generate controlled surface-normal electric-spin textures and measure the associated polarization-sensitive surface responses within a single field of view.

Plasmonic near fields may offer an alternative route in which the local polarization is determined by the structured electromagnetic mode itself. A single planar SPP intrinsically carries transverse spin because of its evanescent character~\cite{Bliokh12PRA,Aiello15NatP,Yanan18ACSN,Qin20PR}. When its in-plane energy flow is spatially structured, the resulting gradients can generate additional transverse-spin components, including one normal to the interface~\cite{Yanan19ACSP,Shi21PNAS,Ichiji23PRA,Kihara25OC}.
This behavior provides a basis for three-dimensional optical-spin textures at metal surfaces, whose spin vectors can form topological patterns such as plasmonic skyrmions and merons~\cite{Du19NatPhys,Yanan20Nat,Lei21PRL,Ghosh21APR}.
However, producing such structured SPP fields has generally required couplers that impose carefully designed phases~\cite{Minovich11PRL,Lin12PRL} or spatially or spatiotemporally shaped incident light~\cite{Ichiji25NCom,Zhang26NCom}. This complexity limits the experimental settings in which these approaches can be used, making a simple configuration for controlling surface-normal spin textures desirable.

In this Letter, we investigate an SPP interference field formed between parallel slits under oblique $s$-polarized plane-wave illumination [Fig.~\ref{Fig:concept}(c)]. The resulting field carries alternating surface-normal electric spin, with a spatial period tunable by the incidence angle through momentum conservation.
FDTD simulations reveal qualitatively different spin textures under forward (air-side) and backward (substrate-side) incidence [Fig.~\ref{Fig:concept}(d)].
Through FDTD field decomposition and simplified modeling, we show that this contrast originates from interference between the SPP and excitation fields, which becomes dominant in the present sine-SPP excitation geometry.

\begin{figure}[t!]
\centering
\includegraphics[width=0.5\textwidth]{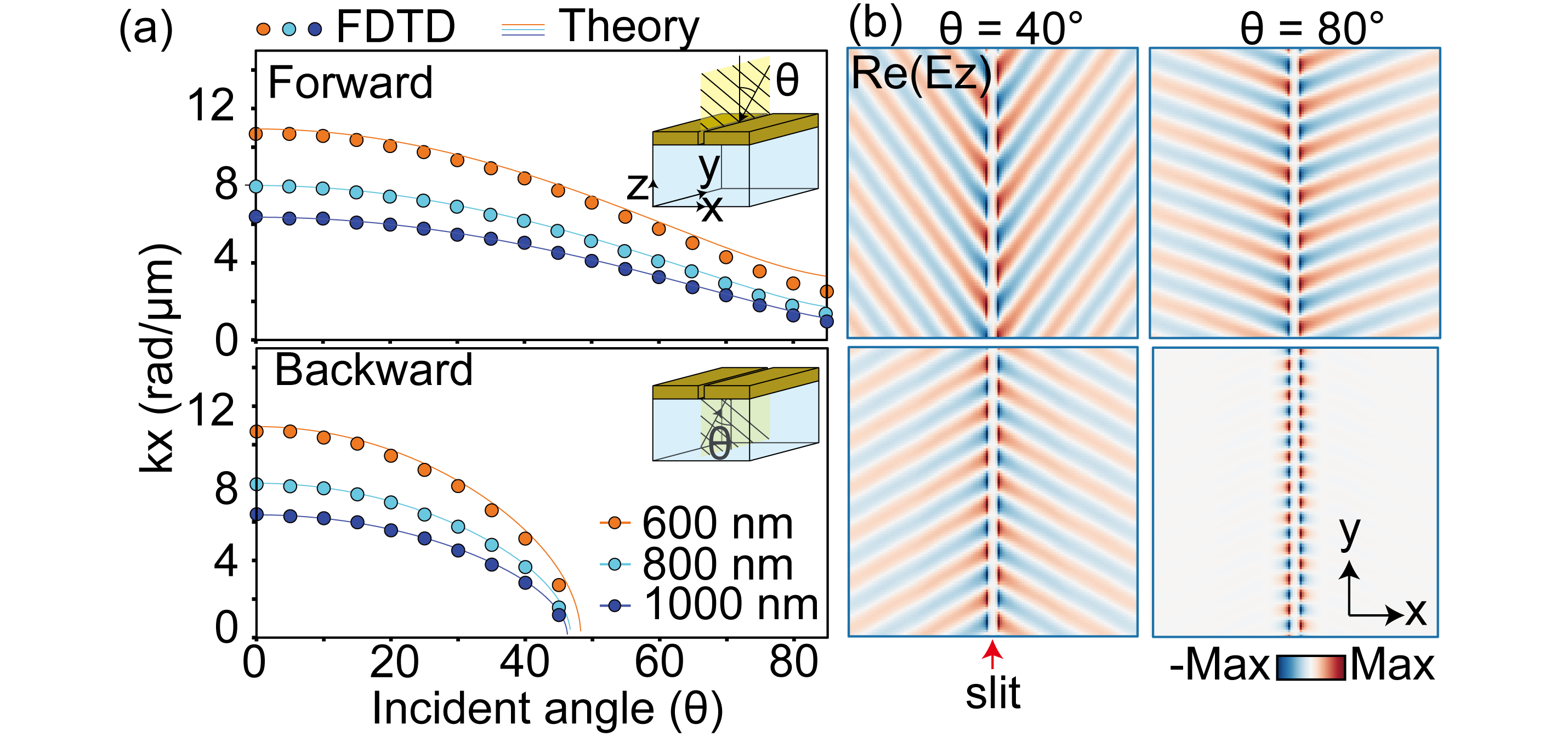}
\caption{SPPs launched from a single slit under oblique $s$-polarized illumination. (a) Slit-normal SPP wave number $k_x$ versus incidence angle $\theta$ for forward (upper panel) and backward (lower panel) incidence. Orange, cyan, and blue indicate wavelengths of 600, 800, and 1000~nm, respectively. Solid curves show the theoretical values calculated from Eq.~(2), and circles show the values extracted from FDTD simulations. (b) Representative spatial distributions of $\mathrm{Re}(E_z)$ at $\lambda=800$~nm for $\theta=40^\circ$ and $80^\circ$.}
\label{Fig}
\end{figure}
First, we characterize the SPP launched by a single slit under oblique illumination. Here, we consider a metal surface in the $xy$ plane containing a slit that extends infinitely along $y$. An $s$-polarized plane wave, with its electric field polarized along $x$, is incident in the $yz$ plane at an angle $\theta$ measured from the surface normal. When the incident wave has a finite wave-vector component along the slit, the launched SPP propagates obliquely with respect to the slit normal~\cite{Genevet15NatN, Nikitin11PRB}. Because the slit is translationally invariant along $y$, phase matching between the incident light and the launched SPP requires conservation of the along-slit wave-vector component. In direct analogy with Snell's law for optical refraction, the propagation angle of the launched SPP is uniquely determined by the incidence angle and the wave numbers of the incident light and the SPP~\cite{Ichiji24JOSAA}. The SPP wave-vector components are therefore given by
\begin{align}
k_y&=k_y^{\mathrm{inc}}=n_{\mathrm{inc}}k_0\sin\theta,\nonumber\\
k_x&=\sqrt{k_{\mathrm{spp}}^2-k_y^2}
\simeq\sqrt{\operatorname{Re}(k_{\mathrm{spp}})^2-k_y^2},
\end{align}
where $k_0=2\pi/\lambda$.

We evaluated the slit-normal SPP wave number $k_x$ using FDTD simulations performed with Tidy3D. 
A 100-nm-thick Au film containing a slit infinitely long along $y$ and 200~nm wide along $x$ was placed on a dielectric substrate with a refractive index of 1.4.
Calculations were performed at wavelengths of 600, 800, and 1000~nm for both air-side (forward) and substrate-side (backward) incidence. The wave number $k_x$ was extracted from the Fourier spectrum of $E_z$. Further numerical details are provided in Appendix A.

Figure~\ref{Fig}(a) compares the $k_x$ values extracted from the FDTD simulations with the theoretical curves calculated using the real-part approximation in Eq.~(2).
The apparent cutoff at $\theta\simeq45^\circ$ under backward incidence can be understood from the momentum-matching condition. Unlike the Kretschmann configuration, momentum matching in the present slit geometry is one-dimensional, involving only the slit-parallel wave-vector component. The momentum-matching condition can therefore be satisfied when \(k_y\lesssim\mathrm{Re}(k_{\mathrm{spp}})\), since oblique SPP propagation allows its slit-parallel wave-vector component, \(k_y^{\mathrm{SPP}}\simeq\sqrt{\mathrm{Re}(k_{\mathrm{spp}})^2-k_x^2}\), to be reduced to match the incident $k_y$.
For $k_y>\mathrm{Re}(k_{\mathrm{spp}})$, the momentum-matching condition cannot be satisfied, resulting in the apparent cutoff. In contrast, under forward incidence, $k_y\leq k_0<\mathrm{Re}(k_{\mathrm{spp}})$, allowing momentum matching at any incidence angle without a corresponding cutoff.
Although the real-part approximation becomes less accurate near the cutoff, the theoretical curves agree reasonably well with the FDTD results at all three wavelengths for both incidence configurations, indicating that the approximation provides a reasonable estimate of $k_x$ under the present conditions.

We next analyze the surface-normal electric spin of the SPP interference field formed between adjacent slits in a periodic array. For this calculation, 200-nm-wide slits were periodically arranged along $x$ with a period of $L\!=\!$ $5.0~\mu$m, allowing the SPPs launched from neighboring slits to overlap and interfere within each interslit region. As shown in Fig.~\ref{Fig}(b), the $E_z$ components of the SPPs launched on opposite sides of a single slit have opposite phases.
Because neighboring slits are driven in phase, their inward-propagating components interfere destructively at the midpoint $x=0$, producing an out-of-plane field with a central node and a spatial dependence $E_z\propto\sin(k_xx)$. In analogy with cosine-type SPP interference fields~\cite{Lin12PRL,Shi21PNAS}, we hereafter refer to this interference field as a sine SPP. The resonance order $n$ is defined as the number of antinodes between adjacent slits, giving an approximate resonance condition $k_x L \simeq n\pi$.

Figures~3(a)--3(c) show the incidence-angle dependence of the total electric-field amplitude $|\mathbf{E}|$, the circular-component imbalance $\Delta_\sigma=|E_+|^2-|E_-|^2$,
and the normalized parameter $s_3=\Delta_\sigma/(|E_x|^2+|E_y|^2)$ at $y=0$, respectively.
The discrete sine-SPP resonances at $n=4$, 6, 8, 10, and 12 in Fig.~3(a) agree well with the resonance angles predicted from Eq.~(2) and the resonance condition, indicated by the light-blue dashed lines. The $|\mathbf{E}|$ distributions exhibit similar resonance behavior under both incidence conditions, with different resonance angles as expected from Fig.~2.

In contrast, the $\Delta_\sigma$ and $s_3$ distributions in Figs.~3(b,c) differ qualitatively between the two incidence configurations. To illustrate these differences more clearly, we plot the spatial cross sections at the $n=12$ resonance in the bottom panels of Figs.~3(a)--3(c).
The spatial period of $\Delta_\sigma$ under forward incidence is approximately twice that under backward incidence. By comparing $|\mathbf{E}|$ and $\Delta_\sigma$, we can confirm that the spin extrema are spatially shifted by approximately $\pi/4$ in $k_xx$ relative to the field-intensity maxima under backward incidence, whereas they coincide under forward incidence.
This contrast is also evident in the normalized parameter $s_3$, whose spatial distribution closely resembles that of $\Delta_\sigma$ under forward incidence but differs substantially under backward incidence.

\begin{figure}[t!]
\centering
\includegraphics[width=0.5\textwidth]{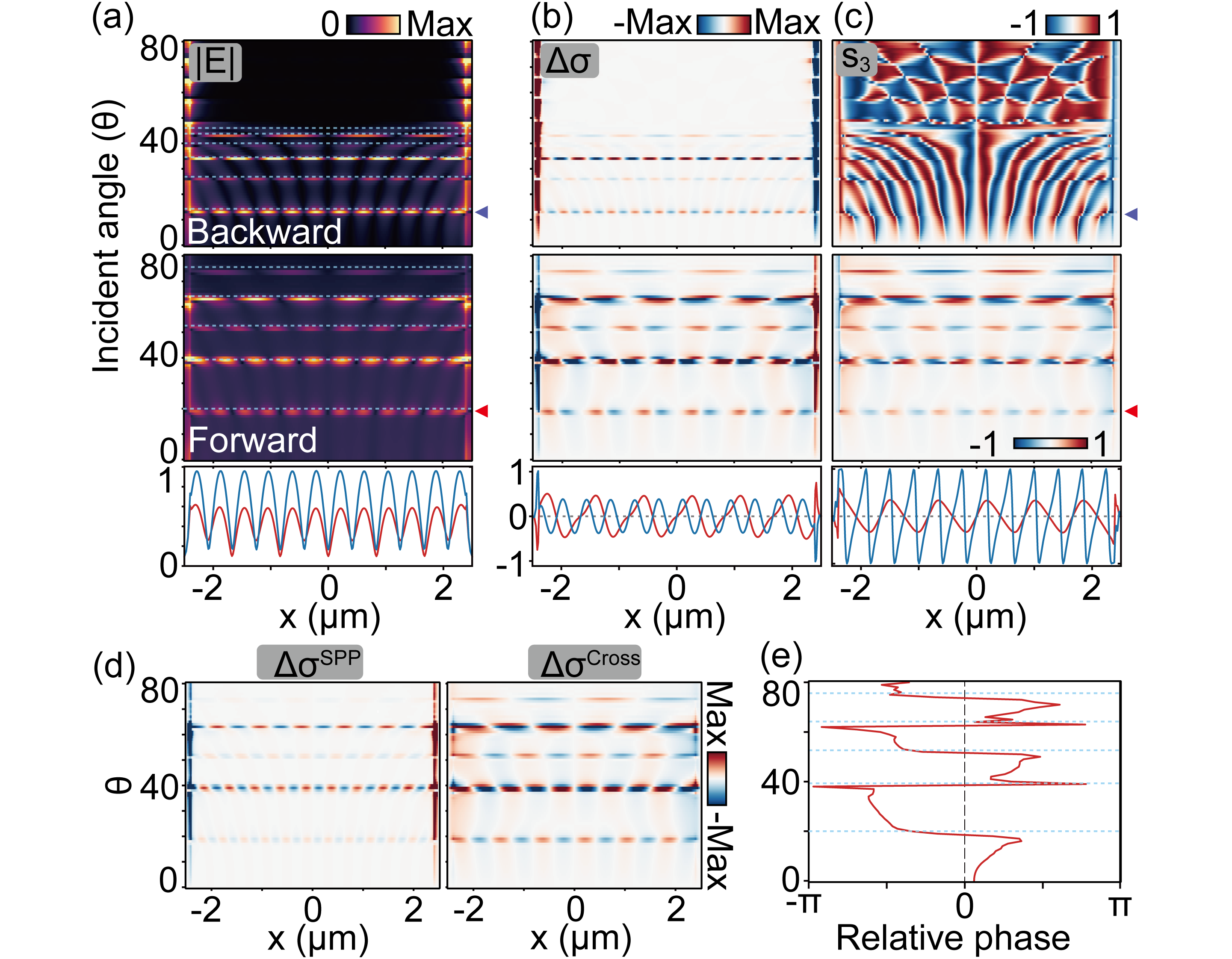}
\caption{Angle-dependent sine-SPP field and spin textures at $\lambda=800$~nm for a 5.0-$\mu$m slit period. (a) Field amplitude $|\mathbf{E}|$ at $y=0$. (b) Circular-component imbalance $\Delta_\sigma=|E_+|^2-|E_-|^2$. (c) Normalized parameter $s_3=\Delta_\sigma/(|E_x|^2+|E_y|^2)$. The upper and middle panels show backward and forward
incidence, respectively, while the bottom panels show spatial cross sections at the $n=12$ resonances. Blue and red lines show the backward and forward incidence, respectively.
The light-blue dashed lines in (a) indicate the predicted resonance angles.
(d) Contributions of the SPP term $\Delta_\sigma^{\mathrm{SPP}}$ (left) and the cross term
$\Delta_\sigma^{\mathrm{cross}}$ (right) under forward incidence. (e) Relative phase $\Delta\phi=\arg(E_z^{\mathrm{slit}})-\arg(E_x^{\mathrm{no\ slit}})$ at $x=0.1~\mu$m and $y=0$. }
\label{Fig:SineTextures}
\end{figure}

These differences can be explained by the different contributions of the excitation field. At the observed surface, the electromagnetic field is dominated by the SPP contribution under backward incidence, whereas under forward incidence it is formed by interference between the SPP and excitation fields~\cite{Zhang11PRB}.
The simpler case is backward incidence, where the direct excitation field is effectively absent at the observed surface. The obliquely launched SPPs retain a common time-averaged energy flow $P_y(x)$ along the slits, so the resulting sine SPP is not a strict standing wave and the spin-flow relation established for cosine SPPs applies~\cite{Shi21PNAS,Ichiji23PRA}.
For a structured evanescent surface wave in a homogeneous lossless isotropic medium, the transverse SAM and time-averaged Poynting vector are related by $\mathbf{S}=(2\omega^2)^{-1}\nabla\times\mathbf{P}$, and the corresponding electric-field spin texture follows the same spatial relation for conventional interference fields~\cite{Nori15PRX,Ichiji26PRA}.
Accordingly, the backward-incidence profiles in Fig.~3(a-c), where opposite-signed extrema of $\Delta_\sigma$ occur near the steepest slopes on either side of each $|\mathbf{E}|$ maximum, are consistent with this relation.

Under forward incidence, by contrast, interference with the excitation light modifies this sine SPP spin texture. Because the SPP near field undergoes a resonance-induced phase evolution whereas the excitation field remains nonresonant, their relative phase changes across each resonance and alters the local in-plane electric-field rotation~\cite{Cao17PRB,Ichiji25NL}.
The $s$-polarized excitation contributes only to $E_x$ at the sample surface, whereas the obliquely launched SPP field contains both $E_x$ and $E_y$ components. Consequently, the circular-component imbalance can be expressed as
\begin{align}
\Delta_\sigma
&=
-2\operatorname{Im}\!\left[
\left(E_x^{\mathrm{SPP}}+E_x^{\mathrm{light}}\right)^*E_y^{\mathrm{SPP}}
\right]
\\
&\equiv
\Delta_\sigma^{\mathrm{SPP}}
+\Delta_\sigma^{\mathrm{cross}}.
\end{align}
This decomposition also explains the contrasting $s_3$ distributions in Fig.~3(c).
Under backward incidence, the total field is dominated by the SPP, so $s_3$ reflects the SPP polarization texture even where the field intensity is weak. Under forward incidence, the excitation field provides a spatially uniform, nonresonant, linearly polarized component.
Consequently, $s_3$ is pronounced where the SPP field is strongly enhanced but suppressed elsewhere, resulting in a distribution similar to that of $\Delta_\sigma$.

To evaluate the contributions in Eq.~(4), we used FDTD simulations with and without the slits, taking the latter as the excitation-field reference. We defined
$E_x^{\mathrm{SPP}}=E_x^{\mathrm{slit}}-E_x^{\mathrm{no\ slit}}$,
$E_x^{\mathrm{light}}=E_x^{\mathrm{no\ slit}}$, and
$E_y^{\mathrm{SPP}}=E_y^{\mathrm{slit}}$,
and calculated $\Delta_\sigma^{\mathrm{SPP}}$ and $\Delta_\sigma^{\mathrm{cross}}$ using Eq.~(4).
Figure~3(d) shows the contributions of $\Delta_\sigma^{\mathrm{SPP}}$ and
$\Delta_\sigma^{\mathrm{cross}}$, under forward incidence. The SPP contribution exhibits a derivative-like spatial profile similar to that observed under backward incidence, whereas the cross term has approximately twice the spatial period and reproduces the characteristic spin texture under forward incidence.
These results confirm that the cross term plays a dominant role in forming the distinctive spin texture under forward incidence.

Furthermore, $\Delta_\sigma^{\mathrm{cross}}$ changes sign across some of the resonances, most notably near the $n=10$ resonance at $\theta\simeq40^\circ$ and the
$n=6$ resonance at $\theta\simeq63^\circ$. Corresponding sign reversals are also observed in the total $\Delta_\sigma$ in Fig.~3(b). This behavior can be explained by the resonance-induced phase shift of the SPP field relative to the excitation field.
Figure~3(e) shows the relative phase between the sine SPP and the excitation field,
$\Delta\phi=\arg(E_z^{\mathrm{slit}})-\arg(E_x^{\mathrm{no\ slit}})$, evaluated at $x=0.1~\mu$m and $y=0$.
The abrupt phase shifts observed around the resonance angles are characteristic of a Lorentz-oscillator-like resonant response and are consistent with the sign reversals of $\Delta_\sigma$ in Fig.~3(b).
 
\begin{figure}[t!]
\centering
\includegraphics[width=0.5\textwidth]{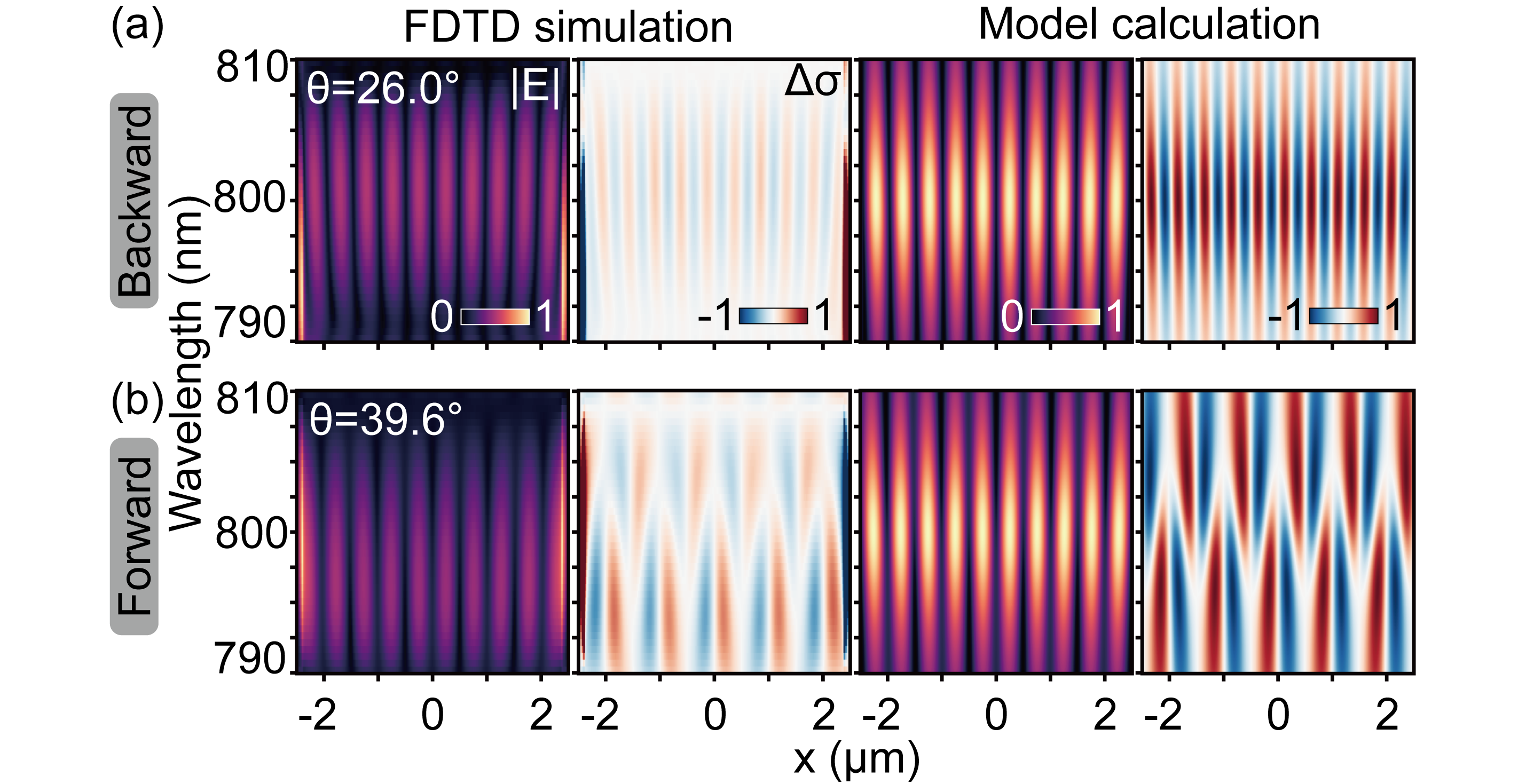}
\caption{Wavelength dependence of the field amplitude and circular component imbalance for backward and forward incidence. (a) Wavelength-resolved distributions of $|\mathbf{E}|$ and $\Delta_\sigma$ under backward incidence at $\theta=26.0^\circ$. (b) Corresponding distributions under forward incidence at $\theta=39.6^\circ$. Within each pair, $|\mathbf{E}|$ and $\Delta_\sigma$ are shown from left to right. Each map is normalized by its maximum absolute value.}
\label{Fig:angle}
\end{figure}

To demonstrate the role of the resonance-induced phase shift in the sign reversals of $\Delta_\sigma$, we constructed simplified models of the sine SPP under backward and forward incidence.
Following the field construction used for cosine SPPs~\cite{Shi21PNAS,Ichiji23PRA}, we modeled the spectral SPP amplitude using a Lorentz oscillator:
\begin{equation}
A_{\mathrm{spp}}(\omega)
=
A_0
\frac{\gamma\omega_{\mathrm{r}}}
{\omega_{\mathrm{r}}^2-\omega^2-i\gamma\omega},
\end{equation}
where $\omega_{\mathrm r}$ and $\gamma$ are the resonance frequency and damping rate, respectively.
Under backward incidence, the total field consists only of the SPP field, whereas under forward incidence, a spatially uniform excitation field $E_x^{\mathrm{light}}=E_0$ is added.
Taking ($\omega_{\mathrm r}$) to correspond to a wavelength of 800 nm, ($\gamma/\omega_{\mathrm r}$) and ($A_0/E_0$) were set to 0.015 and 5, respectively. Further details of the model are shown in Appendix B.

Figure~4 compares wavelength-resolved FDTD simulations at fixed incidence angles with the corresponding model calculations. The model qualitatively reproduces the FDTD distributions for both incidence configurations in Fig.~4(a,b).
In the forward-incidence case, $\Delta_\sigma$ reverses sign within the spectral bandwidth of a 100-fs pulse centered at 800~nm. The agreement between the model and FDTD results confirms that interference with the excitation field accounts for the qualitative difference between backward and forward incidence.

In summary, we have shown that oblique \(s\)-polarized illumination of parallel slits produces a sine SPP with an alternating circular-component imbalance corresponding to surface-normal electric spin.
The angle-selected SPP wave vector determines the spatial period, and the sine-SPP interference is accompanied by a periodic spin texture.
Under backward incidence, the resulting spin distribution is consistent with the conventional spin-flow relation, whereas under forward incidence, interference with the excitation field produces a distinct spin texture.

Compared with conventional $p$-polarized slit excitation used to generate cosine-type SPP fields~\cite{Lin12PRL,Ichiji25NCom}, where the dominant in-plane components of the excitation and SPP fields are parallel, the present oblique $s$-polarized geometry allows their orthogonal in-plane components to generate a dominant cross term in the surface-normal electric spin under forward incidence.
In addition, our analysis of how the cross term between the SPP near field and excitation light affects spin textures may also be extended to more complex systems, such as
plasmonic skyrmion lattices~\cite{Lei21PRL,Ghosh21APR}.

Although our simulations assume infinitely long slits, obliquely propagating SPPs have been experimentally excited using sufficiently long, FIB-milled slits illuminated with a finite-sized beam~\cite{Ichiji25NCom}. Such a simple configuration is advantageous for experimental implementation. The parallel-slit geometry also provides access to large slit-normal wave-vector components $k_x$ that are difficult to realize in conventional cosine-SPP configurations~\cite{Lin12PRL,Shi21PNAS}. 
A practical limitation is that the primary control parameter is the incidence angle.
In experimental configurations with a fixed illumination geometry, $k_y$ is fixed by the prescribed incidence angle. To obtain the resonant sine-SPP interference considered here, the slit spacing must therefore be designed for that angle.
Nevertheless, the constraint on $k_x$ may be relaxed by engineering the SPP dispersion.
For example, a thin dielectric overlayer modifies $k_{\mathrm{spp}}$ through its refractive index and thickness, allowing $k_x$ to be tuned without changing the illumination
geometry~\cite{Pockrand78SS,Ichiji22Nano,Kamada26NL}.

The formation of periodic regions of opposite $\Delta_\sigma$ within a single field of view is particularly advantageous for microscopy, because polarization-sensitive responses to opposite local handedness can be compared under identical illumination and detection conditions. The present configuration therefore provides a simple route to spatially resolved studies of polarization-sensitive surface responses under oblique illumination.

\section*{Acknowledgments} This work was supported by JSPS KAKENHI (JP25K17906) and Fujikura Foundation. The author thanks A. Kubo for invaluable discussions and support, and T. Fukuda for helpful comments.
\newpage

\appendix
\renewcommand\thefigure{S\arabic{figure}} 
\setcounter{figure}{0}  

\section{FDTD Simulation Conditions}
The simulations were performed using Tidy3D. A 100-nm-thick Au film containing 200-nm-wide slits was placed on a substrate with a refractive index of 1.4, with air above the film. The Au permittivity was taken from the Olmon2012evaporated model in the Tidy3D material library~\cite{Olmon}.
An s-polarized plane wave illuminated the film from either the air or substrate side.
Bloch-periodic and periodic boundary conditions were applied along the slit direction ($y$) and slit-normal direction ($x$), respectively, with PMLs along $z$. The computational domain extended 1.4 $\mu$m above and below the film, and the source was placed 0.75 $\mu$m from the corresponding Au interface. The cell length along $y$ was fixed at 2.0 $\mu$m, while the period along $x$ depended on the calculation. A Gaussian pulse with central frequency ($f_0=c/\lambda_0$) and frequency width ($f_0/50$) was used, with a maximum simulation time of 2.5 ps. The complex electric-field components were recorded 10 nm above the Au surface. 
An automatically generated nonuniform mesh was used, with a target resolution of 10 grid cells per wavelength in each material, resulting in a minimum grid spacing of approximately 15 nm. Refining it to approximately 10 nm changed the extracted ($k_x$) by 1.8$\%$ at 600 nm and 60$^\circ$ under forward incidence. Considering the finite wave-number resolution of the Fourier analysis and the computational cost, we adopted the original mesh for the simulations.
For the single-slit calculations in Fig. 2, the period along ($x$) was set to 50 $\mu$m to approximate an isolated slit, beyond which the extracted ($k_x$) was insensitive to further increases. Calculations were performed at wavelengths of 600, 800, and 1000 nm for both air-side (forward) and substrate-side (backward) incidence. The slit-normal wave number ($k_x$) was extracted from the dominant Fourier peak of the ($E_z$) cross section at ($y=0$). For Figs. 3 and 4, the slit-normal period was reduced to 5.0 $\mu$m. Figure 3 was obtained by sweeping the incidence angle and evaluating the fields at 800 nm. Reference calculations without slits used otherwise identical settings for the field decomposition.
The source setting and in-plane boundary conditions were changed for Fig. 4. Figures 2 and 3 used the default FixedInPlaneKSpec setting, which fixes the incident in-plane wave vector across the spectrum, with the prescribed incidence angle applying only at the central frequency. For Fig. 4, FixedAngleSpec and periodic boundaries along both $x$ and $y$ were used to maintain a constant incidence angle across the spectrum. Wavelength-resolved fields over 790–810 nm were obtained from broadband simulations at 26.0$^\circ$ for backward incidence and 39.6$^\circ$ for forward incidence.

\section{Simplified Model Calculation}
Following the field construction used for structured SPPs~\cite{Shi21PNAS, Ichiji23PRA}, we modeled the sine-SPP field as a superposition of two SPPs with opposite slit-normal wave-vector components. The SPP wave number was calculated using the same Au permittivity as in the FDTD simulations:
\begin{equation*}
k_{\mathrm{spp}} = k_0\sqrt{\frac{\varepsilon_{\mathrm{Au}}\varepsilon_d}{\varepsilon_{\mathrm{Au}}+\varepsilon_d}},
\qquad
K = \operatorname{Re}(k_{\mathrm{spp}}).
\end{equation*}
where ($k_0=2\pi/\lambda$) and ($\varepsilon_{d}=1$). The real-part approximation was used to construct the spatial field distribution, neglecting SPP propagation loss along the surface. At each wavelength, the wave-vector components and the evanescent decay constant in air were calculated as
\begin{align*}
k_{y}&=n_{inc}k_{0}\sin\theta, \\
k_x&=\sqrt{(K^2-k_y^2)}, \\
\kappa&=\sqrt{(K^2-\varepsilon_{d}k_0^2)}.
\end{align*}
Here, $n_{inc}=1.0$ and 1.4 for forward and backward incidence, respectively.
Using the time dependence ($e^{-i\omega t}$), the spectral SPP amplitude was described by a Lorentz response,
\begin{equation*}
A_{\mathrm{SPP}}(\omega) =
A_0\frac{\gamma\omega_r}
{\omega_r^2-\omega^2-i\gamma\omega},
\end{equation*}

where $A_0$ is a real constant, $\omega_r$ is the resonance frequency, and $\gamma$ is the damping constant. At the observation plane, the electric-field components were constructed as
\begin{align*}
E_z^{\mathrm{SPP}}
&= A_{\mathrm{SPP}}\sin(k_x x)e^{ik_y y},\\
E_x^{\mathrm{SPP}}
&= -\frac{\kappa k_x}{K^2}
A_{\mathrm{SPP}}\cos(k_x x)e^{ik_y y},\\
E_y^{\mathrm{SPP}}
&= -i\frac{\kappa k_y}{K^2}
A_{\mathrm{SPP}}\sin(k_x x)e^{ik_y y}.
\end{align*}

Under backward incidence, the total field was taken to be the SPP field alone. Under forward incidence, a nonresonant excitation field polarized along $x$ was added:
\begin{align*}
E_x&=E_x^{SPP}+E_0 e^{ik_{y}y},\\
E_y&=E_y^{SPP},\\
E_z&=E_z^{SPP},
\end{align*}
where $E_0$ is real and independent of wavelength. The field amplitude and circular-component imbalance were then evaluated as ($|E|=\sqrt{(|E_x |^2+|E_y |^2+|E_z |^2)}$) and ($\Delta_\sigma=-2\operatorname{Im}(E_x^*E_y)$).
For Fig. 4, the resonance wavelength was set to 800 nm, with ($\gamma/\omega_r=0.015$) and ($A_0/E_0=5$). The incidence angles were 26.0$^\circ$ for backward incidence and 39.6$^\circ$ for forward incidence. The fields were evaluated at ($y = 0$) over ($-2.5\leq x\leq2.5$) $\mu$m and wavelengths of 790–810 nm.

\bibliography{sorsamp}

@article{Bliokh12PRA,
  Author = {Bliokh, Konstantin Y. and Nori, Franco},
  Title = {Transverse spin of a surface polariton},
  Journal = {Phys. Rev. A},
  Year = {2012},
  Volume = {85},
  Number = {6},
  Pages = {061801(R)},
}

@article{Aiello15NatP,
  Author = {Aiello, Andrea and Banzer, Peter and Neugebauer, Martin and Leuchs, Gerd},
  Title = {From transverse angular momentum to photonic wheels},
  Journal = {Nat. Photonics},
  Year = {2015},
  Volume = {9},
  Number = {12},
  Pages = {789--795},
}

@article{Shi21PNAS,
  Author = {Shi, Peng and Du, Luping and Li, Congcong and Zayats, Anatoly V. and Yuan, Xiaocong},
  Title = {Transverse spin dynamics in structured electromagnetic guided waves},
  Journal = {Proc. Natl. Acad. Sci. U.S.A.},
  Year = {2021},
  Volume = {118},
  Number = {6},
  Pages = {e2018816118},
}

@article{Lei21PRL,
  Author = {Lei, Xinrui and Yang, Aiping and Shi, Peng and Xie, Zhenwei and Du, Luping and Zayats, Anatoly V. and Yuan, Xiaocong},
  Title = {Photonic spin lattices: symmetry constraints for skyrmion and meron topologies},
  Journal = {Phys. Rev. Lett.},
  Year = {2021},
  Volume = {127},
  Number = {23},
  Pages = {237403},
}

@article{Ghosh21APR,
Author = {Ghosh, Atreyie and Yang, Sena and Dai, Yanan and Zhou, Zhikang and Wang,
   Tianyi and Huang, Chen-Bin and Petek, Hrvoje},
Title = {A topological lattice of plasmonic merons},
Journal = {Appl. Phys. Rev.},
Year = {2021},
Volume = {8},
Number = {4},
pages = {041413},
}

@article{Lin12PRL,
  Author = {Lin, Jiao and Dellinger, Jean and Genevet, Patrice and Cluzel, Beno{\^i}t and de Fornel, Fr{\'e}d{\'e}rique and Capasso, Federico},
  Title = {Cosine-{Gauss} plasmon beam: a localized long-range nondiffracting surface wave},
  Journal = {Phys. Rev. Lett.},
  Year = {2012},
  Volume = {109},
  Number = {9},
  Pages = {093904},
}

@article{Ichiji25NCom,
  Author = {Ichiji, Naoki and Kikuchi, Hibiki and Yessenov, Murat and Schepler, Kenneth L. and Abouraddy, Ayman F. and Kubo, Atsushi},
  Title = {Observation of space-time surface plasmon polaritons},
  Journal = {Nat. Commun.},
  Year = {2025},
  Volume = {16},
  Number = {1},
  Pages = {10697},
}

@article{Genevet15NatN,
  Author = {Genevet, Patrice and Wintz, Daniel and Ambrosio, Antonio and She, Alan and Blanchard, Romain and Capasso, Federico},
  Title = {Controlled steering of {Cherenkov} surface plasmon wakes with a one-dimensional metamaterial},
  Journal = {Nat. Nanotechnol.},
  Year = {2015},
  Volume = {10},
  Number = {9},
  Pages = {804--809},
}

@article{Ichiji23PRA,
  Author = {Ichiji, Naoki and Oue, Daigo and Yessenov, Murat and Schepler, Kenneth L. and Abouraddy, Ayman F. and Kubo, Atsushi},
  Title = {Transverse spin angular momentum of a space-time surface plasmon polariton wave packet},
  Journal = {Phys. Rev. A},
  Year = {2023},
  Volume = {107},
  Number = {6},
  Pages = {063517},
}

@article{Pockrand78SS,
  Author = {Pockrand, I.},
  Title = {Surface plasma oscillations at silver surfaces with thin transparent and absorbing coatings},
  Journal = {Surf. Sci.},
  Year = {1978},
  Volume = {72},
  Number = {3},
  Pages = {577--588},
}

@article{Ichiji22Nano,
  Author = {Ichiji, Naoki and Otake, Yuka and Kubo, Atsushi},
  Title = {Femtosecond imaging of spatial deformation of surface plasmon polariton wave packet during resonant interaction with nanocavity},
  Journal = {Nanophotonics},
  Year = {2022},
  Volume = {11},
  Number = {7},
  Pages = {1321--1333},
}

@article{Word17Ultra,
Author = {Word, Robert C. and K{\"o}nenkamp, Rolf},
Title = {Photonic and plasmonic surface field distributions characterized with normal- and oblique-incidence multi-photon {PEEM}},
Journal = {Ultramicroscopy},
Year = {2017},
Volume = {183},
Pages = {43--48},
doi = {10.1016/j.ultramic.2017.05.012},
}

@article{Forbes24JO,
Author = {Forbes, Kayn A.},
Title = {Spin angular momentum and optical chirality of Poincar{\'e} vector vortex beams},
Journal = {J. Opt.},
Year = {2024},
Volume = {26},
Number = {12},
pages = {125401},
doi = {10.1088/2040-8986/ad8cec},
}

@article{Ishito23NatP,
  Author = {Ishito, Kyosuke and Mao, Huiling and Kousaka, Yusuke and Togawa, Yoshihiko and Iwasaki, Satoshi and Zhang, Tiantian and Murakami, Shuichi and Kishine, Jun-ichiro and Satoh, Takuya},
  Title = {Truly chiral phonons in {$\alpha$-HgS}},
  Journal = {Nat. Phys.},
  Year = {2023},
  Volume = {19},
  Number = {1},
  Pages = {35--39},
}

@article{Mak12NatN,
  Author = {Mak, Kin Fai and He, Keliang and Shan, Jie and Heinz, Tony F.},
  Title = {Control of valley polarization in monolayer {MoS$_2$} by optical helicity},
  Journal = {Nat. Nanotechnol.},
  Year = {2012},
  Volume = {7},
  Number = {8},
  Pages = {494--498},
}

@article{Choi17NatCom,
  Author = {Choi, Gyung-Min and Schleife, Andr{\'e} and Cahill, David G.},
  Title = {Optical-helicity-driven magnetization dynamics in metallic ferromagnets},
  Journal = {Nat. Commun.},
  Year = {2017},
  Volume = {8},
  Number = {1},
  Pages = {15085},
}

@article{Yanan18ACSN,
  Author = {Dai, Yanan and Dabrowski, Maciej and Apkarian, Vartkess A. and Petek, Hrvoje},
  Title = {Ultrafast microscopy of spin-momentum-locked surface plasmon polaritons},
  Journal = {ACS Nano},
  Year = {2018},
  Volume = {12},
  Number = {7},
  Pages = {6588--6596},
}

@article{Yanan19ACSP,
  Author = {Dai, Yanan and Petek, Hrvoje},
  Title = {Plasmonic spin-{Hall} effect in surface plasmon polariton focusing},
  Journal = {ACS Photon.},
  Year = {2019},
  Volume = {6},
  Number = {8},
  Pages = {2005--2013},
}

@article{Qin20PR,
  Author = {Qin, Yulu and Ji, Boyu and Song, Xiaowei and Lin, Jingquan},
  Title = {Disclosing transverse spin angular momentum of surface plasmon polaritons through independent spatiotemporal imaging of its in-plane and out-of-plane electric field components},
  Journal = {Photonics Res.},
  Year = {2020},
  Volume = {8},
  Number = {6},
  Pages = {1042--1048},
}

@article{Yanan20Nat,
  Author = {Dai, Yanan and Zhou, Zhikang and Ghosh, Atreyie and Mong, Roger S. K. and Kubo, Atsushi and Huang, Chen-Bin and Petek, Hrvoje},
  Title = {Plasmonic topological quasiparticle on the nanometre and femtosecond scales},
  Journal = {Nature},
  Year = {2020},
  Volume = {588},
  Number = {7839},
  Pages = {616--619},
}

@article{Kamada26NL,
    author = {Kamada, Kazuki and Kim, DaeGwi and Shibuta, Masahiro},
    title = {Visualization of Internal Plasmonic Wave Photosensitized by Quantum Dots},
    journal = {Nano Letters},
    volume = {26},
    number = {23},
    pages = {7808-7815},
    year = {2026},
}

@article{Ichiji26PRA,
  title = {Transverse spin texture in optical non-Hermitian skin modes},
  author = {Ichiji, Naoki and Takeda, Issei and Yoda, Taiki and Moritake, Yuto and Notomi, Masaya and Ashihara, Satoshi},
  journal = {Phys. Rev. A},
  volume = {114},
  issue = {3},
  pages = {033522},
  numpages = {13},
  year = {2026},
}

@article{Minovich11PRL,
  Author = {Minovich, Alexander and Klein, Angela E. and Janunts, Norik and Pertsch, Thomas and Neshev, Dragomir N. and Kivshar, Yuri S.},
  Title = {Generation and near-field imaging of {Airy} surface plasmons},
  Journal = {Phys. Rev. Lett.},
  Year = {2011},
  Volume = {107},
  Number = {11},
  Pages = {116802},
}

@article{Zhang26NCom,
  Author = {Zhang, Yuquan and Ma, Haixiang and Ju, Zhendong and Chen, Xusheng and Chen, Yixuan and Xie, Xi and Du, Luping and Min, Changjun and Yuan, Xiaocong},
  Title = {Structureless excitation and manipulation of dynamic holographic plasmonic slides},
  Journal = {Nat. Commun.},
  Year = {2026},
  Volume = {17},
  Number = {1},
  Pages = {2946},
}

@article{Neugebauer18PRX,
  Author = {Neugebauer, Martin and Eismann, J{\"o}rg S. and Bauer, Thomas and Banzer, Peter},
  Title = {Magnetic and electric transverse spin density of spatially confined light},
  Journal = {Phys. Rev. X},
  Year = {2018},
  Volume = {8},
  Number = {2},
  Pages = {021042},
}

@article{Kihara25OC,
  Author = {Kihara, Kotaro and Motoi, Kei and Ichiji, Naoki and Kubo, Atsushi},
  Title = {Spin angular momentum of femtosecond striped space-time surface plasmon polaritons},
  Journal = {Opt. Commun.},
  Year = {2025},
  Volume = {593},
  Pages = {132217},
}

@article{Zhang11PRB,
  Author = {Zhang, Lingxiao and Kubo, Atsushi and Wang, Leiming and Petek, Hrvoje and Seideman, Tamar},
  Title = {Imaging of surface plasmon polariton fields excited at a nanometer-scale slit},
  Journal = {Phys. Rev. B},
  Year = {2011},
  Volume = {84},
  Number = {24},
  Pages = {245442},
}

@article{Cao17PRB,
  Author = {Cao, Z. L. and Yiu, L. Y. and Zhang, Z. Q. and Chan, C. T. and Ong, H. C.},
  Title = {Understanding the role of surface plasmon polaritons in two-dimensional achiral nanohole arrays for polarization conversion},
  Journal = {Phys. Rev. B},
  Year = {2017},
  Volume = {95},
  Number = {15},
  Pages = {155415},
}

@article{Ichiji25NL,
  Author = {Ichiji, Naoki and Ishida, Takuya and Morichika, Ikki and Oue, Daigo and Tatsuma, Tetsu and Ashihara, Satoshi},
  Title = {Selective enhancement of the optical chirality and spin angular momentum in plasmonic near-fields},
  Journal = {Nano Lett.},
  Year = {2025},
  Volume = {25},
  Number = {33},
  Pages = {12578--12584},
}

@article{Du19NatPhys,
    Author = {Du, Luping and Yang, Aiping and Zayats, Anatoly V. and Yuan, Xiaocong},
    Title = {Deep-subwavelength features of photonic skyrmions in a confined electromagnetic field with orbital angular momentum},
    Journal = {Nat. Phys.},
    Year = {2019},
    Volume = {15},
    pages = {650-654},
}

@ARTICLE{Nori15PRX,
  AUTHOR =       {Bekshaev, Aleksandr Y. and Bliokh, Konstantin Y. and Nori, Franco},
  TITLE =        {Transverse spin and momentum in two-wave interference},
  JOURNAL =      {Phys. Rev. X},
  YEAR =         {2015},
  volume =       {5},
  pages =        {011039},
}

@article{Olmon,
   author = {Olmon, Robert L. and Slovick, Brian and Johnson, Timothy W. and Shelton, David and Oh, Sang-Hyun and Boreman, Glenn D. and Raschke, Markus B.},
   title = {Optical dielectric function of gold},
   journal = {Phys. Rev. B},
   volume = {86},
   year = {2012},
}

@article{Nikitin11PRB,
  title = {Oblique launching of optical surface waves by a subwavelength slit},
  author = {Nikitin, A. Yu. and Garc\'{\i}a-Vidal, F. J. and Mart\'{\i}n-Moreno, L.},
  journal = {Phys. Rev. B},
  volume = {83},
  issue = {15},
  pages = {155448},
  numpages = {7},
  year = {2011},
}

@ARTICLE{Ichiji24JOSAA,
  AUTHOR =       {N. Ichiji and M. Yessenov and K. L. Schepler and A. F. Abouraddy and A. Kubo},
  TITLE =        {Exciting space-time surface plasmon polaritons by irradiating a nanoslit structure},
  JOURNAL =      {J. Opt. Soc. Am. A},
  YEAR =         {2024},
  volume =       {41},
  pages =        {396-405},
}
\end{document}